\documentclass[twocolumn]{aastex61}

\received{XXX, 2016}
\revised{XXX, 2016}
\accepted{\today}

\shorttitle{Is HESS J1809-193 a pevatron?}
\shortauthors{Araya, M.}

\begin{document}

\title{GeV emission in the region of HESS J1809-193 and HESS J1813-178: is HESS J1809-193 a proton pevatron?}

\correspondingauthor{Miguel Araya}
\email{miguel.araya@ucr.ac.cr}

\author{Miguel Araya}
\affiliation{Escuela de F\'isica \& Centro de Investigaciones Espaciales (CINESPA),\\
Universidad de Costa Rica, San Pedro de Montes de Oca, 2060 San Jos\'e,\\
Costa Rica}

\begin{abstract}
HESS J1809-193 is an unidentified TeV source discovered by the High Energy Stereoscopic System and originally classified as a pulsar wind nebula (PWN) candidate associated with the pulsar PSR J1809-1917. However, a recent study of deep radio observations and the interstellar medium near the source has found evidence for a hadronic scenario for the gamma-rays. Here, a detailed study of the GeV emission in the region using data from the \emph{Fermi} LAT is presented. The GeV emission has an extended morphology in the region of the TeV emission and the overall spectrum can be accounted for by a cosmic ray population having a simple power-law spectrum with energies extending up to 1 PeV. However, the spectrum at tens of TeV should be observed more deeply in the future to confirm its hadronic nature, and other scenarios involving combinations of leptonic and hadronic emission from several of the known supernova remnants in the region cannot be ruled out. The nearby TeV source HESS J1813-178, thought to be a PWN, is also studied in detail at GeV energies and we find a region of significant emission which is much more extended than the TeV emission and whose spectrum is softer than expected from a PWN but similar to those seen in several star forming regions that are believed to accelerate protons. There is marginal evidence for a GeV point source at the location of the X-ray PWN, beside the extended emission.
\end{abstract}


\section{Introduction} \label{sec:intro}
Although supernova remnants (SNRs) have for a long time been considered the main sources of Galactic cosmic rays and even though evidence for the acceleration of hadrons in some of these systems has been found \citep{2013Sci...339..807A,2016ApJ...816..100J}, the indirectly measured energies reached by the particles in SNRs are not even close to the estimated highest energies in Galactic cosmic rays, of the order of PeV. A source of PeV cosmic rays was recently discovered close to the Galactic center \citep{2016Natur.531..476H} but its nature has not been understood. In fact, many sources of TeV gamma-rays remain unidentified \citep[e.g.,][]{2015ICRC...34..773D,2017ApJ...843...40A} and multiwavelngth studies of their environment could help reveal the nature of some of them and contribute to solve the puzzle of the cosmic ray origin.

We carried out a detailed study of GeV emission in the region of the TeV sources HESS J1809-193 and HESS J1813-178 using data from the \emph{Fermi} Large Area Telescope (LAT), a converter/tracker telescope that detects photons in the energy range between 20 MeV and $> 500$ MeV \citep{2009ApJ...697.1071A}. The LAT has revealed thousands of GeV sources and has expanded our knowledge of the high-energy universe \citep[e.g.,][]{2010ApJS..188..405A,2015ApJS..218...23A,2017ApJS..232...18A}.

HESS J1809-193 was discovered during a search for pulsar wind nebulae (PWNe) carried out with the High Energy Stereoscopic System (H.E.S.S) \citep{2007A&A...472..489A}. The brightness profile was fit with an elliptical Gaussian with widths $32'\pm4'$ and $15'\pm2'$. The center of the ellipse is at the location RA=$18^{\mbox{\tiny h}}10^{\mbox{\tiny m}}31^{\mbox{\tiny s}}\pm 12^{\mbox{\tiny s}}$, Dec=$-19\degr 18'\pm 2'$ (J2000) \citep{2007A&A...472..489A}. The TeV spectrum of the source was fit by a power-law with a spectral index of $2.2\pm0.1_{\mbox{\tiny stat}}$. Later, a preliminary analysis using more data found a spectral index of $2.23\pm 0.05_{\mbox{\tiny stat}}$, but a 2D symmetric Gaussian was chosen to fit the region of brightest emission \citep{2008AIPC.1085..285R}. The latest TeV maps of the Galactic plane by H.E.S.S. also seem to show a source with an elongated morphology \citep{2015ICRC...34..773D}.

The region of HESS J1809-193 is complicated. It contains several SNRs, G11.0-0.0 being coincident with the peak of the TeV emission. In addition, a pulsar with a high-spin down energy loss rate, PSR J1809-1917, is located around 0.1$\degr$ from this peak. Since diffuse X-ray emission thought to belong to the PWN associated with PSR J1809-1917 has been found \citep{2003ApJ...589..253B,2007ApJ...670..655K}, a PWN origin for HESS J1809-193 was for some time preferred. Subsequent X-ray observations with Suzaku \citep{2010PASJ...62..179A} confirmed the existence of diffuse hard X-ray emission which was previously reported by ASCA and \emph{Chandra}, and showed that the diffuse emission extends for at least 21 pc (at a distance of 3.5 kpc to the pulsar PSR J1809-1917). No signs of cooling via synchrotron emission along this distance were seen, contrary to expectations for a young PWN, and thus requiring perhaps small magnetic turbulence to allow for such rapid electron diffusion \citep{2010PASJ...62..179A}. This scenario has been suggested to occur in older SNR or PWN such as HESS J1825-137, with an associated pulsar having a characteristic age of 21 kyr \citep[e.g.,][]{2005ApJ...621..793B,2009PASJ...61S.189U}. For comparison, the characteristic age of PSR J1809-1917 is 51 kyr \citep{2002MNRAS.335..275M}, while the age of the SNR G11.0-0.0 is unknown. There is also a second energetic pulsar, PSR J1811-1925, farther from the center of HESS J1809-193 and more distant than J1809-1917 \citep[5 kpc,][]{2005AJ....129.1993M}, at the center of SNR G11.2-0.3. \cite{2014ApJ...796...34R} argued that this pulsar and the SNR as well as the X-ray binary XTE J1810–189 and binary candidate Suzaku J1811–1900 found to the NE of the peak of TeV emission cannot be responsible for the gamma-rays.

\cite{2016A&A...587A..71C} recently carried out a search for the counterpart to the proposed PWN driven by PSR J1809-1917 using deep radio observations from the Karl G. Jansky Very large Array, but no such counterpart was found. Additionally, their study of the interstellar medium in the direction of the source revealed several molecular clouds at the shock front of the SNR G11.0-0.0, a location that coincides with the brightest peak of the TeV gamma-rays, as well as evidence for interaction between the SNR and the clouds. The distance to the SNR and the clouds estimated by \cite{2016A&A...587A..71C} is 3 kpc. In view of this, hadronic interactions were proposed by these authors to explain the origin of the gamma-ray emission.

Two unidentified LAT point sources have been associated to HESS J1809-193 or found in the region of the TeV source, 3FGL J1810.1-1910 and 3FGL J1811.3-1927c \citep{2015ApJS..218...23A}. Very close to 3FGL J1811.3-1927c the source 3FHL J1811.5-1927, found in The Third Catalog of Hard Fermi-LAT Sources \citep{2017ApJS..232...18A}, is associated to the pulsar SWIFT J1811.5-1925.

Another interesting source seen near HESS J1809-193 is HESS J1813-178. It was discovered in 2005 by H.E.S.S. \citep{2005Sci...307.1938A}. It is a compact TeV source (i.e., nearly pointlike for H.E.S.S.). When the 2D brightness profile was fit by the function $\rho \propto \mathrm{e}^{-r^2/2\sigma^2}$, an extension value of $\sigma=3'$ (with an statistical error of 10-30\%) was found, and a peak position at RA=$18^{\mbox{\tiny h}}13^{\mbox{\tiny m}}36^{\mbox{\tiny s}}$, Dec=$-17\degr 50'$ (J2000), with an statistical error in the range 1-2 arcmin \citep{2005Sci...307.1938A}. No counterpart to the gamma-rays was known originally but the source has been associated with the SNR G12.82-0.02 after the discovery of non-thermal radio and X-ray emissions in the region \citep{2005ApJ...629L.105B,2005ApJ...629L.109U}. The source properties were later confirmed at TeV energies by MAGIC \citep{2006ApJ...637L..41A}, which reported a spectrum consistent with a simple power-law with a spectral index of $2.1\pm 0.2_{\mbox{\tiny stat}}\pm 0.2_{\mbox{\tiny sys}}$ (in the range 0.4-10 TeV). \emph{XMM-Newton} X-ray and NANTEN observations later revealed a resolved non-thermal X-ray source in the center of the SNR and a giant molecular cloud near HESS J1813-178, respectively \citep{2007A&A...470..249F}. Attributing the gamma-rays to hadrons in the shell of the SNR, a maximum proton energy of 100 TeV is needed, while a leptonic/PWN scenario linking the compact X-ray core and the TeV gamma-rays requires electron energies in excess of 1 PeV \citep{2007A&A...470..249F}.

A highly energetic pulsar, capable of powering the particles producing the TeV emission, PSR J1813-1749, was later discovered near the center of the SNR G12.82-0.02 overlaping the TeV source \citep{2009ApJ...700L.158G}. The distance to the pulsar was deduced to be 4.7 kpc by association with an adjacent young stellar cluster \citep{2009ApJ...700L.158G,2008ApJ...683L.155M}, however, a large X-ray column density measured was found to exceed the visual extinction to the cluster, which means that the actual distance may be larger \citep{2012ApJ...753L..14H}. A search for GeV emission from PWN with the LAT did not yield any detection for HESS J1813-178 recently \citep{2013ApJ...773...77A}, but two unidentified LAT point sources are found in the region, namely, 3FGL J1814.0-1757c and 3FGL J1814.1-1734c \citep{2015ApJS..218...23A}. The source is currently classified as a PWN candidate.

We report the detection of extended emission in the range 0.5-500 GeV in the regions of HESS J1809-193 and HESS J1813-178. In both cases, the emission shows a simple porwer-law spectrum with a similar spectral index as that of the corresponding TeV source ($\sim$2.2 and 2.1, respectively) which can be explained by hadronic processes and is difficult to explain with inverse Compton (IC) emission from high-energy electrons alone. For the case of HESS J1809-193, the hadronic scenario requires a proton distribution extending up to 1 PeV. In the case of HESS J1813-178, the GeV source is much more extended than the TeV emission.

\section{\emph{Fermi} LAT data} \label{sec:LAT}
We analized data from the beginning of the mission (August 2008) to July 2017 with the most recent response functions (effective area, point-spread function and energy resolution, Pass 8) and version v10r0p5 of the ScienceTools. The event class SOURCE was used and events having a maximum zenith angle of 90$\degr$ were selected to avoid contamination from events produced in interactions with the Earth, as well as time intervals when the data quality was good. The data are binned spatially using a scale of 0.1$\degr$ per pixel and, for exposure calculation, in ten logarithmically spaced bins per decade in energy. Filtered events reconstructed within a radius of 20$\degr$ around the reported TeV centroid of HESS J1809-193, and in the energy range 0.2-500 GeV, were used for analysis.

The LAT tools fit a model to the data (including the residual charged particles and diffuse gamma-rays) making use of the maximum likelihood technique \citep{1996ApJ...461..396M}. The likelihood $L$ is the probability of obtaining the data given an input model. The model includes the spatial features and locations of sources and their spectra, including the Galactic diffuse emission, given by the file gll\_iem\_v06.fits, and the residual background and extragalactic emission, given by the file iso\_P8R2\_SOURCE\_V6\_v06.txt, which are provided by the LAT team. The likelihood is calculated as the product of the Poisson probabilities for each bin that is used to divide the data. The fit starts with an initial model containing the diffuse emission and the sources found in the LAT Third Source Catalog \citep[3FGL,][]{2015ApJS..218...23A} within the region of interest, but additional sources were added later to improve the model as it might be expected for a larger data set than used in the catalog. The detection significance as well as the best-fit morphological and spectral parameters of sources were estimated with the test statistic (TS), defined as $-2\times$log$(L_0/L)$, with $L$ and $L_0$ the values of the likelihood functions for nested models. The TS is maximized when $L$ is maximized and it has a known distribution, therefore, its value is a measure of the goodness-of-fit. Comparing the TS values for different models in this work provided a criterion to select the best-fit spectral and morphological properties of the emission in the region of the H.E.S.S. sources.

\subsection{Morphology of the sources}
In order to take advantage of the improved resolution at higher energies, we studied the morphology of the sources above 5 GeV. As a first step, we optimized the location and shape of new significant sources seen in the residuals maps. Later, when the analysis was done at lower energies, even more sources (with soft gamma-ray spectra) had to be added to the region of interest to improve the fit and fully consider their effect. Whenever a point source was added to the model, the tool \emph{gtfindsrc} was used to optimize its location. When instead an extended excess was found, a systematic search for the maximum TS value was done by fitting an uniform disc template along a grid of positions and using a range of radii at each position (changing each in steps of 0.1$\degr$). For simplicity, a simple power-law spectrum is used for additional background sources, fitting also the normalization and spectral index each time.

The unidentified 3FGL sources 3FGL J1810.1-1910, 3FGL J1811.3-1927c in the region of HESS J1809-193, as well as 3FGL J1814.0-1757c and 3FGL J1814.1-1734c possibly associated to HESS J1813-178, were removed from the model. They all have a relatively flat spectral energy distribution and thus are probably part of the extended emission found here with a more detailed analysis. The source 3FGL J1808.5-1952, to the SW of HESS J1809-193, remained in the model, as it has a soft GeV spectrum and is associated to a globular cluster and thus is not part of the TeV source. In the fit, the normalizations of the 3FGL sources located within $7\degr$ from the center of the region of interest were set free, while the rest were fixed to the cataloged values. Based on the inspection of the residual analysis obtained after an initial fit with the 3FGL model, it was evident that additional sources from the hard LAT source catalog (3FHL) in the region of interest needed to be added, these included extended sources such as 3FHL J1804.7-2144e (containing the unidentified TeV source HESS J1804-216), 3FHL J1800.5-2343e in the W28 region, and the point source 3FHL J1801.5-2450, classified as a pulsar in the W28 region \citep{2017ApJS..232...18A}. Both the normalization and spectral index of these 3FHL sources were kept free in the fits.

\begin{figure}[ht!]
\includegraphics[width=9.5cm,height=8cm]{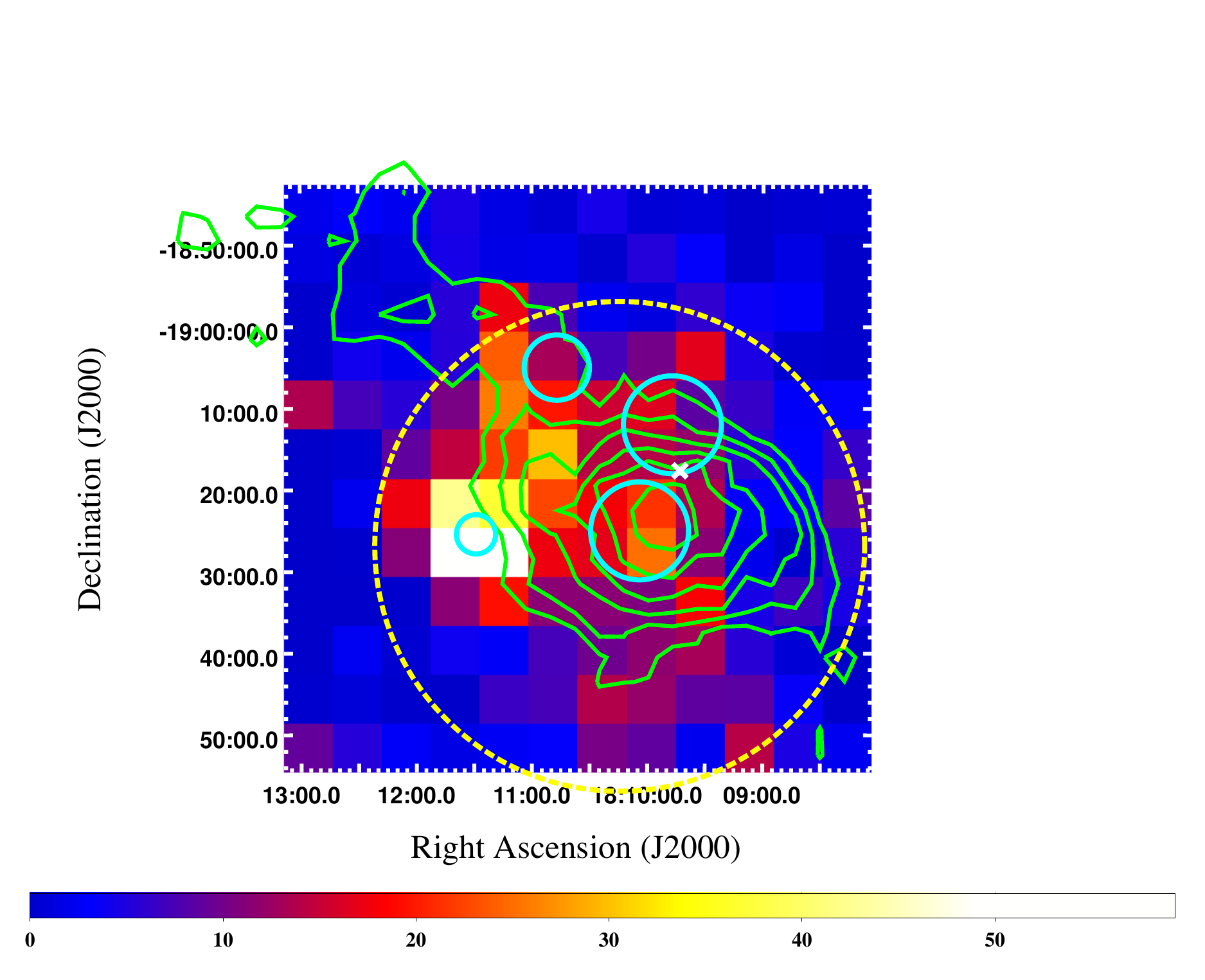}
\caption{TS map for a point source hypothesis in the region of HESS J1809-193. It was obtained with LAT data above 5 GeV after fitting a point source located at each pixel in the map. The position of the pulsar PSR J1809-1917 is indicated by a white `x'. The dashed yellow circle represents the border of the best fit disc template found in this work and the green contours represent the H.E.S.S. significance contours at levels from 3$\sigma$ to 8$\sigma$ in steps of 1$\sigma$, from \cite{2007A&A...472..489A}. The smaller cyan circles represent the sizes and locations of SNRs in the region according to the database of SNR observations by \cite{2012AdSpR..49.1313F} (see also http://www.physics.umanitoba.ca/snr/SNRcat/), clockwise starting from the one located at the highest declination: G11.4-0.1, G11.1+0.1, G11.0-0.0 and G11.2-0.3.}\label{fig1}
\end{figure}

After improving the model in this way for events above 5 GeV, the new residuals show apparently extended excess emission in the regions of HESS J1809-193 and HESS J1813-178. The GeV excess at the location of HESS J1809-193 seemed to be elongated, approximately following the TeV emission. Fig. \ref{fig1} shows a TS map obtained for HESS J1809-193 using a point source hypotesis which reveals significant emission.

 A detailed study of the morphologies was carried out and the results are shown in Table \ref{tab1}. We used different hypotheses for the morphology of the emission, including the point sources reported in the 3FGL catalog for HESS J1809-193. In the case of HESS J1813-178, the point sources from the 3FGL catalog are located relatively far away from the peak of the excess and thus new locations were optimized with the larger data set used here. For HESS J1809-193, the TeV template obtained from the count map shown in the H.E.S.S. discovery paper \cite{2007A&A...472..489A} was also used.

For both H.E.S.S. sources, considerable residuals are seen after subtracting the point source hypothesis model from the data. As can be seen in Table \ref{tab1}, the quantity TS$_{\mbox{\tiny disc}}-$TS$_{\mbox{\tiny point}}$=$2\times$(log($L_{\mbox{\tiny disc}}$)-log($L_{\mbox{\tiny point}}$)), a measure of the source extension, is much greater than 16, the typical threshold used to claim that a source is extended \citep[e.g.,][]{2016ApJS..224....8A}. Additionally, in both cases, the TS of the disc morphology is greater than the TS of the two point sources, which is another criterion that has been applied to distinguish between an extended source and a combination of point sources \citep{2017arXiv170200476T}. The best-fit disc radii, spectral indices and center locations (J2000) for HESS J1809-193 and HESS J1813-178 are $0.5\pm0.15\degr$, $2.22\pm0.12$, RA=$18^{\mbox{\tiny h}}10^{\mbox{\tiny m}}22^{\mbox{\tiny s}}$, Dec=$-19\degr 26'$ ($\pm 0.2\degr$), and $0.6\pm0.06\degr$, $2.07\pm0.09$, RA=$18^{\mbox{\tiny h}}13^{\mbox{\tiny m}}10^{\mbox{\tiny s}}$, Dec=$-17\degr 37'$ ($\pm 0.05\degr$), respectively. The radii and location errors are given at the 3$\sigma$ level.

After subtracting the best-fit model with the uniform disc hypothesis for HESS J1813-178 to the data some photons are observed clustered at the location of the TeV peak. We fit the position of a new point source on top of the disc template for HESS J1813-178, and as can be seen in the Table there is some evidence at the $3\sigma$-level for the contribution of a point source besides the extended emission (TS$_{\mbox{\tiny disc+point}}-$TS$_{\mbox{\tiny disc}}$=18, for 4 additional degrees of freedom). The location of this point source was found to be RA=$18^{\mbox{\tiny h}}13^{\mbox{\tiny m}}51^{\mbox{\tiny s}}$, Dec=$-17\degr 51'$ (and $3'$ 1$\sigma$ error circle radius), which is consistent with the peak of the TeV emission. However, the individual TS of the point source obtained in a fit was slightly below the threshold of 25 to claim a detection. This was also true in the other energy intervals studied and we decided to only use a disc in the final model for HESS J1813-178.

For HESS J1809-193 and HESS J1813-178 the extended templates provide a much better fit compared to the point source hypothesis. In the case of HESS J1809-193, both the disc morphology and TeV counts map template can be considered good representations of the GeV morphology. We used the disc template to estimate the spectral parameters and used the TeV counts map template to estimate a systematic uncertainty that we added to the statistical uncertainty for those parameters. We conclude that extended emission is significantly detected in the region of the TeV sources.

\begin{table*}
	\centering
	\caption{TS values above 5 GeV for different spatial morphologies with respect to the null hypothesis.}
	\begin{tabular}{|c|c|c|c|}
          \hline
	  \hline
	  \textbf{Spatial model} & \textbf{HESS J1809-193} & \textbf{HESS J1813-178} & \textbf{D.o.f.$^{a}$} \\
          & & &\\
          \hline
	  Uniform disc  & 97.5  & 162 & 5 \\
          \hline
          TeV counts map  &  76.7 & ---  & 2  \\
          \hline
          1 point source & 44 & 65 & 4 (2)$^{b}$ \\
          \hline
          2 point sources  & 60 & 92  & 8 (4)$^{b}$ \\
          \hline
          Disc + point source & --- & 180 & 9 \\
          \hline
	\end{tabular}\\
        \textsuperscript{$a$}\footnotesize{Additional degrees of freedom with respect to the null hypothesis.}\\
        \textsuperscript{$b$}\footnotesize{The point sources in the HESS J1809-193 region are located at the known positions of the 3FGL sources, and thus the model has fewer degrees of freedom in this case.}\\
        \label{tab1}
\end{table*}

\subsection{Spectral shapes and additional soft background sources}
When we fit the model found previously to data in the energy interval 1-500 GeV, several new residuals unrelated to HESS J1809-193 and HESS J1813-178 appeared in different parts of the region of interest, most notably at the location of the unidentified TeV source HESS J1808-204, and close to the point source 3FGL J1809.2-2016c. The emission seems to be extended and we optimized the location and size of a disc template to account for it. The center position and radius of the resulting disc were RA=$18^{\mbox{\tiny h}}08^{\mbox{\tiny m}}23^{\mbox{\tiny s}}$, Dec=$-20\degr 30'$ and $0.3\degr$, respectively. The source significance and spectral index were $8\sigma$ and $3.1\pm0.7$ above 1 GeV. Similarly, we added other sources that are relatively significant and have very soft spectra (with spectral indices in the range 2.6-3.1) at the coordinates (all given in degrees) RA,Dec: 273.53,-17.8 to the East of HESS J1813-178, 271.65,-21.345 close to W30, 274.492, -20.079 outside the Galactic plane and 275.08, -16.21 about 2.2$\degr$ NE of HESS J1813-178. Finally, a 0.3$\degr$-radius disc template was found to the South of the region of interest, with center at the coordinates RA=272.96, Dec=-24.106 (with spectral index $\sim$2.8 and overall significance $\sim11\sigma$), having the unidentified source 3FGL J1810.8-2412 at the border, both of which might be related to SNR G7.5-1.7. The addition of all these sources was found to have no effect on the morphological studies done at higher energies.

We chose the energy range 0.5-500 GeV to make the final spectral analysis because substantial soft emission not accounted for by our model was seen below 500 MeV near HESS J1809-193, and that complicated the analysis at the lowest energies. The residuals in the final energy range chosen were acceptable and therefore the model was a good representation of the data. As shown in Table \ref{tab2}, the spectra of the GeV sources coincident with HESS J1809-193 and HESS J1813-178 are better described by a simple power-law with fewer degrees of freedom than the other models corresponding to ``curved'' spectra. The spectral indices and integrated fluxes are, respectively for HESS J1809-193 and HESS J1813-178,  $2.22\pm 0.08$, $(1.68\pm0.1)\times 10^{-8}$ photons cm$^{-2}$ s$^{-1}$ and $2.14\pm 0.04$, $(2.0\pm0.11)\times 10^{-8}$ photons cm$^{-2}$ s$^{-1}$. Table \ref{tab2} shows the final TS values, which correspond to detection significances of 19.5$\sigma$ and 21.6$\sigma$ for HESS J1809-193 and HESS J1813-178, respectively.

\begin{table*}
	\centering
	\caption{TS values$^{a}$ for different spectral shapes for HESS J1809-193 and HESS J1813-178 (0.5-500 GeV).}
	\begin{tabular}{|c|c|c|}
          \hline
	  \hline
	  \textbf{Spectral shape} & \textbf{HESS J1809-193} & \textbf{HESS J1813-178} \\
          & & \\
          \hline
	  Simple power-law  & 381.5 & 467 \\
          \hline
          LogParabola  & 381.5 & 467 \\
          \hline
          Power-law with & 381.5 & 467 \\
          exponential cutoff &  &  \\
          \hline
	\end{tabular}\\
        \textsuperscript{$a$}\footnotesize{Calculated with respect to the null hypothesis.}\\
        \label{tab2}
\end{table*}

We binned the LAT data in ten energy bins equally spaced logarithmically in the 0.5-500 GeV range to measure the source flux and plot spectral points next to the global fit. Whithin each interval, we fixed the spectral indices of the sources associated to HESS J1809-193 and HESS J1813-178 to the value found above and fit the normalization to calculate the photon flux. 95\%-confidence level upper limits on the flux are calculated for intervals without a source detection. The individual flux values are consistent with the fits in the entire energy range and we present the spectral energy distribution (SED) models in the following section.

\section{Modeling}
\subsection{HESS J1809-193}
\begin{figure}[ht!]
\includegraphics[width=9cm,height=5.5cm]{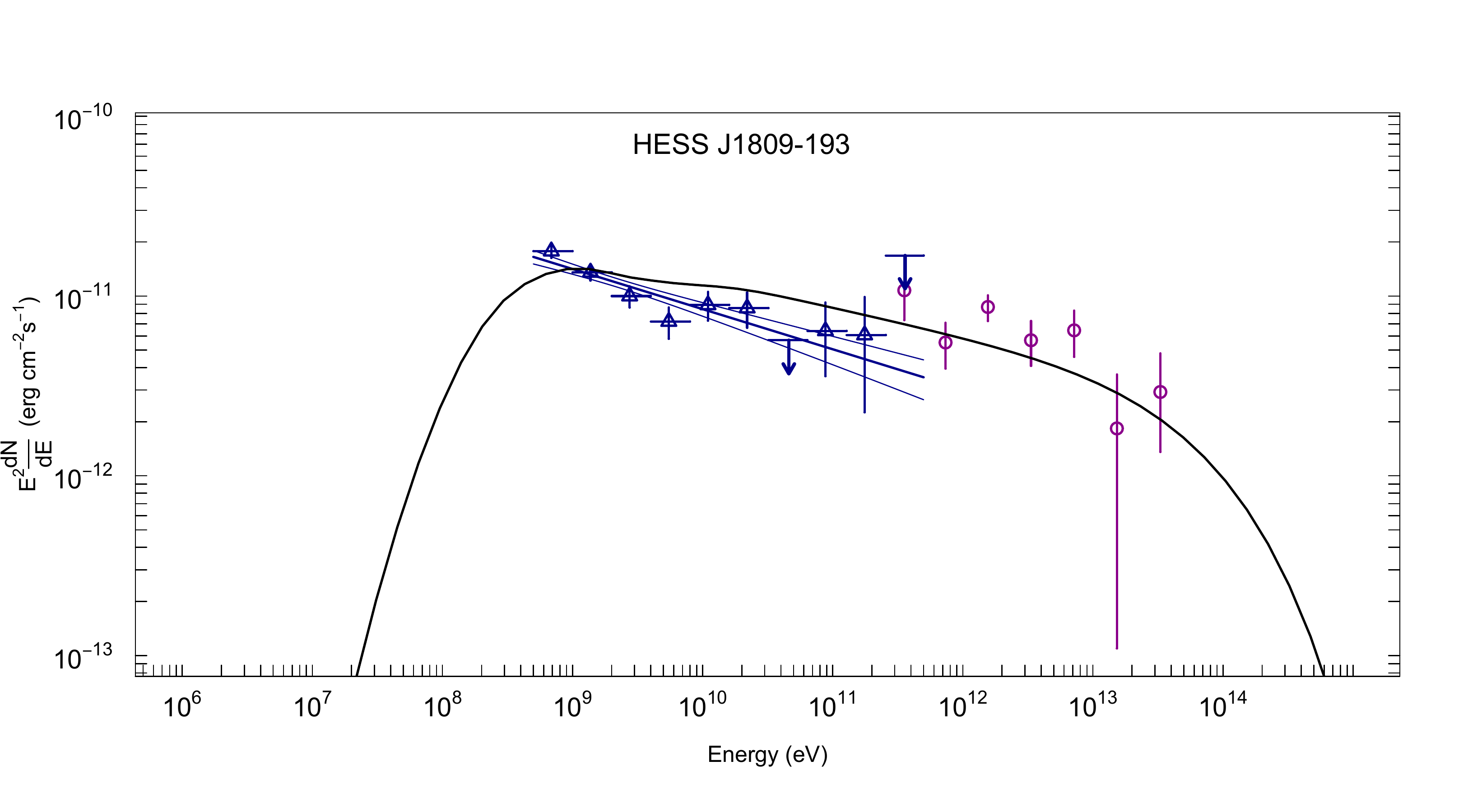}
\caption{Hadronic model for HESS J1809-193. The particle distribution is a power-law with an exponential cutoff (index 2.22 and cutoff energy 1 PeV). The triangles are the LAT data from this work (with the overall 0.5-500 GeV best-fit and uncertainty band) and the circles are H.E.S.S. measurements \citep{2007A&A...472..489A}.}\label{fig2}
\end{figure}

The gamma-ray spectrum of the emission in the region of HESS J1809-193 is extended and shows a spectral index of $2.22\pm 0.05$ which is entirely consistent with that measured at TeV energies, $2.23\pm 0.05$ \citep{2008AIPC.1085..285R}. Motivated by this fact, we fit the GeV-TeV spectral points to a radiative model from hadronic interactions through Markov Chain Monte Carlo sampling of likelihood distributions of the fit parameters (with a simple power-law particle spectrum) using the numerical code \emph{Naima} \citep{2015arXiv150903319Z}. This code implements a parametrization of the gamma-ray production cross-sections for \emph{pp} interactions from \cite{2014PhRvD..90l3014K}. The resulting best-fit particle spectral index and its 1$\sigma$ error was $2.22\pm 0.02$ ($\chi^2=27$ for 14 degrees of freedom). This result is not surprising since the gamma-ray spectrum follows the particle distribution and both of their spectral indices should be similar. Using a particle distribution that is a power-law with an exponential cutoff, the resulting cutoff energy in the fit tends towards the upper bound in the allowed range when the range is increased, and thus the data were not able to constrain it. The data and model are shown in Fig. \ref{fig2} for a particle distribution with a cutoff energy of 1 PeV. It should be kept in mind that the simplified model shown here using one single population may be an oversimplification of the real situation. A complete explanation of the morphology of the gamma-ray emission should involve the modeling of cosmic ray diffusion, which was not taken into account here.

The hadronic scenario for HESS J1809-193 is supported by the recent observations of \cite{2016A&A...587A..71C} and this work proves that this source could be a cosmic ray pevatron. From the same fit using \emph{Naima} \citep{2015arXiv150903319Z} it was found that the total energy needed in the high-energy protons is $$6.2\times 10^{49}\,\,\mbox{erg} \,\left(\frac{10\,\mbox{cm}^{-3}}{n}\right) \,\left(\frac{d}{3\,\,\mbox{kpc}}\right)^2$$ for a source distance $d$ (in kpc) and a target density $n$ (in cm$^{-3}$).

Typical kinetic energies in SNRs are of the order of $10^{51}$ erg. It is believed that around 10\% of this energy can be transferred to cosmic rays \citep[although this value is highly uncertain, see, e.g.,][]{2013A&A...553A..34D} and therefore using this conversion efficiency an average density of 6 cm$^{-3}$ is required for a source distance of 3 kpc. As mentioned before, \cite{2016A&A...587A..71C} made a thorough analysis of ambient structures in the region of the TeV source. They discovered a system of molecular clouds physically related to the shock front of the SNR G11.0-0.0 at an estimated distance of 3 kpc, coincident with the peak of the TeV emission and with a total proton density in the range $2-3 \times 10^3$ cm$^{-3}$.

The possibility that HESS J1809-193 is accelerating protons to PeV energies is extremely interesting and relevant to the theory of cosmic ray origin. Although the current paradigm for the origin of cosmic rays requires that SNRs act as pevatrons for at least some time during their evolution, no known SNR shows a gamma-ray spectrum that is compatible with this idea. Only one unidentified pevatron candidate is known, and it is located in the Galactic center region \citep{2016Natur.531..476H}. However, a considerable fraction of the TeV emission from this region can alternatively be explained by leptonic emission from a pulsar population expected to exist near the Galactic center \citep{2017arXiv170509293H}. The results presented here do not, of course, prove that HESS J1809-193 is a pevatron but give evidence in favor of a hadronic scenario for the gamma-ray data (specifically at GeV energies). The nature of the GeV to TeV emission has to be confirmed as hadronic in the future (using more detailed multiwavelength observations and modeling) and the spectrum of the source needs to be accurately measured at tens of TeV to confirm the H.E.S.S. results and to explore if the spectrum extends to higher energies than those measured so far. For this purpose, currently operating TeV observatories such as the High-Altitude Water Cherenkov Gamma-Ray Observatory \citep{2013APh....50...26A} may be able to measure the source spectrum at the highest energies, and future TeV observatories such as the Cherenkov Telescope Array \citep{2011NuPhS.212..170H} will offer a substantial improvement in resolution which is crucial to locate the emission regions. The confirmation of the acceleration of cosmic rays to PeV energies in the HESS J1809-193/G11.0-0.0 system would be a considerable advancement in the understanding of the origin of cosmic rays, as it would be the first case involving a known SNR.

\subsubsection{Alternative scenarios: contributions from several SNRs}
The SNRs that are seen in the region have small angular sizes and the emission from their shells would be close to point-like for \emph{Fermi} (see Fig. \ref{fig1}). The gamma-ray emission in the region of HESS J1809-193, on the other hand, is better described by an extended hypothesis. One may explain this emission by proposing the existence of a previously unknown extended SNR in the region to account for the gamma-rays. However, deep radio imaging does not show evidence for such an SNR \citep{2016A&A...587A..71C}, which would be visible at those wavelengths through synchrotron radiation.

Considering the fact that the known SNRs in the region of HESS J1809-193 are located close to each other in the sky it might be possible that the LAT observations cannot distinguish the gamma-ray contributions from them. We point out that the shape of the SED at gamma-ray energies can be explained by leptonic or mixed scenarios involving leptonic and hadronic contributions from several or all of the SNRs in the region. 

The SNRs are likely located at different distances. For G11.4-0.1, \cite{2014SerAJ.189...25P} and \cite{2004AJ....127..355B} give values of 8.4 kpc and 9 kpc, respectively, estimated from the radio surface brightness - diameter ($\Sigma-D$) relation. Some distances found in the literature for the other SNRs are 8.3 kpc for G11.1+0.1 implied by the same relation \citep{2014SerAJ.189...25P}, although if G11.1+0.1 is associated to the pulsar PSR J1809-1917 its distance would be 3.7 kpc \citep{2002MNRAS.335..275M}, and 9.8 kpc and 3.0 kpc for G11.0-0.0 \citep{2014SerAJ.189...25P,2016A&A...587A..71C}. G11.2-0.3 contains a compact X-ray PWN near its center \citep{2003ApJ...588..992R} and it has been studied in detail. Several derived distances or distance constraints for G11.2-0.3 are 10 kpc, 6 kpc, 4.4 kpc and $>$4.5 kpc \citep[][respectively]{2014SerAJ.189...25P,2004AJ....127..355B,2016ApJ...816....1K,1972ApJS...24...49R}. \cite{2004BASI...32..335G} and \cite{2008ApJ...676.1189M} also derived distances to G11.2-0.3 of 4.4 kpc and 5.5-7.7 kpc, respectively.

We may assume that only the sources G11.1+0.1 and G11.0-0.0 are located at comparable distances of 3-4 kpc, and we show next the SED plots with simple leptonic and mixed models that can explain the available data, including the radio fluxes from these two sources. The models shown are not actual fits to the data but simple comparisons of the flux levels assuming the same leptonic spectra is produced in the sources. Although exploring all possible parameter space is outside the scope of this paper, we may give estimates of the values for some of the physical parameters required. Since the distances cited above for G11.4-0.1 were obtained from the $\Sigma-D$ relation whose validity has been questioned strongly in the literature \citep[e.g.,][]{2004BASI...32..335G}, we consider that its distance is unknown and include its fluxes in the plots for comparison.

\begin{figure}[ht!]
\includegraphics[width=9cm,height=5.5cm]{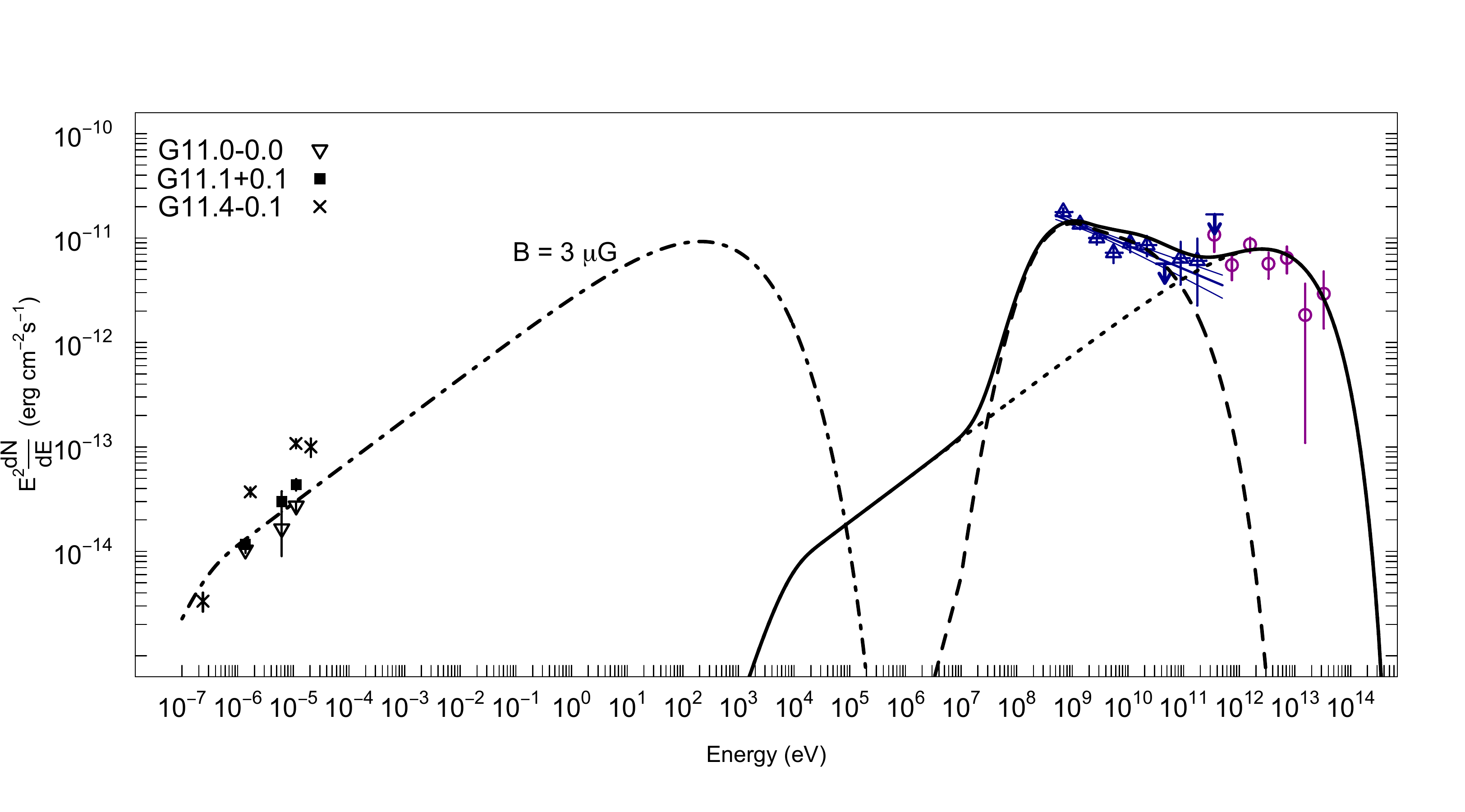}
\caption{The gamma-ray data shown in Fig. \ref{fig2} and the measured synchrotron fluxes. The model is a mixed model involving leptonic (dotted line) and hadronic (dashed line) components for the GeV-TeV emission from HESS J1809-193. The solid line is the sum of these two components. The corresponding fluxes from synchrotron emission are shown for an average magnetic field value of 3 $\mu$G (dash-dotted lines).}\label{fig3}
\end{figure}

Fig. \ref{fig3} shows a mixed leptonic and hadronic model where the TeV emission comes from high energy leptons scattering off cosmic microwave background photons (IC-CMB), and the GeV emission, having a spectrum that is softer than expected from IC-CMB interactions, comes from hadronic collisions. The particle distributions in both cases are described by a power-law with an exponential cutoff and index 2.2. The cutoff energies are 1 TeV for hadrons and 50 TeV for leptons, while their total energies are $4.8\times 10^{49}\,\,\mbox{erg} \,\left(\frac{10\,\mbox{cm}^{-3}}{n}\right)$ and $1.0\times 10^{49}$ erg, respectively, for a source distance of 3 kpc and target density $n$. The existence of high-energy leptons gives rise to synchrotron fluxes at a level which depends on the average magnetic field. Although the magnetic field is expected to be amplified in an SNR, with respect to its value in the interstellar medium \citep[e.g.,][]{2001MNRAS.321..433B}, some observations imply relatively low average field values ($\sim$ 5-10 $\mu$G) for some gamma-ray emitting SNRs \citep[e.g.,][]{2016ApJ...819...98A,2017ApJ...843...12A}. Fig. \ref{fig3} shows the resulting synchrotron fluxes and the measured fluxes of three SNRs, chosen based on the previous discussion. Synchrotron fluxes were obtained for a magnetic field of 3 $\mu$G. The radio data were taken from the literature \citep{1989ApJS...71..799K,2006ApJ...639L..25B}. The particles producing the gamma-rays may thus be supplied by the SNRs in the region and their level of radio emission is compatible with the model. Of course, only observations of X-ray synchrotron emission can provide evidence for leptons having the necessary energies to account for the TeV emission.

\begin{figure}[ht!]
\includegraphics[width=9cm,height=5.5cm]{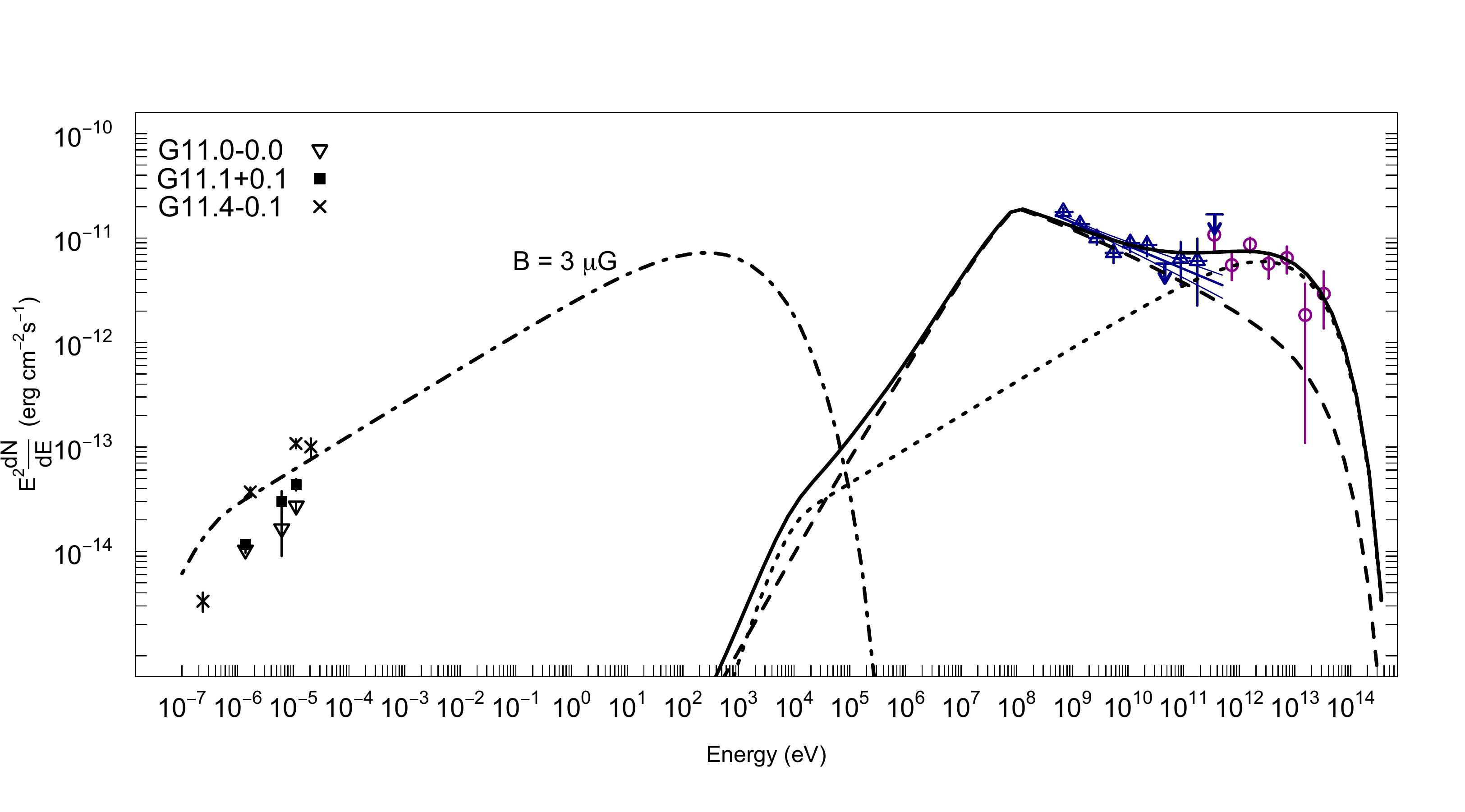}
\caption{The same data shown in Fig. \ref{fig3} with a leptonic model for HESS J1809-193. The dashed line is the bremsstrahlung spectrum and the dotted line is IC-CMB emission, while the solid line is the sum of these two components. Synchrotron emission is shown for a magnetic field of 3 $\mu$G (dash-dotted lines).}\label{fig4}
\end{figure}

The gamma-ray SED of HESS J1809-193 is also compatible with a purely leptonic model consisting of bremsstrahlung emission producing most of the GeV fluxes and IC-CMB producing TeV emission. It was found that a single population with an energy distribution in the form of a power-law with index 2.35 and an exponential cutoff energy of 70 TeV reproduces the observed high-energy fluxes, as shown in Fig. \ref{fig4}. The resulting fluxes from synchrotron emission are compatible with the sum of the individual radio fluxes from G11.1+0.1 and G11.0-0.0. The required total energy in leptons is $5.4\times10^{49}$ erg for a distance of 3 kpc and the density of the target material for bremsstrahlung emission is 10 cm$^{-3}$. The level of X-ray synchrotron emission from the same leptons should be detectable even for a relatively low average magnetic field of 3 $\mu$G.

Finally, as pointed out by \cite{2016A&A...587A..71C}, star formation activity could also play a role in producing the gamma-rays seen from HESS J1809−-193. Several bright HII regions lie within the boundary of the VHE source (see the next section for details on some of the observed gamma-ray spectra and morphologies of star forming regions).

\subsection{HESS J1813-178}
In the case of HESS J1813-178 similar scenarios to those for HESS J1809-193 could be considered. Interestingly, the spectral indices measured at TeV and GeV energies are compatible ($\sim 2.1$). However, the TeV emission seems to be compact while the GeV emission is very extended (as seen in Section \ref{sec:LAT}, the radius found was $0.6\pm0.06\degr$ for a disc-like hypothesis). Different populations of cosmic rays might produce independent components of emission at GeV and TeV energies. However, little is known about the effects of, for example, a changing environment in the spectra of gamma-ray emission produced by hadrons or leptons. Remarkably, for example, similar morphologies have been seen at GeV and TeV energies for SNR IC 443 whose gamma-rays are of hadronic origin \citep{2013Sci...339..807A} but there are different ambient properties throughout the emission regions \citep{2015ICRC...34..875H}.

From the fact that the GeV emission that we found here is very extended while current TeV observations seem to show only a compact source at the highest energies, we only considered here an scenario for the GeV gamma-rays, as the TeV emission might come from a PWN powered by PSR J1813-1749. The spectral shape of the emission can be easily explained by a hadronic accelerator, perhaps the SNR G12.82-0.02 or one of the other SNRs in the region \citep[see a catalogue of SNRs by][]{2012AdSpR..49.1313F}, and hard to explain in a leptonic scenario where the expected emission is harder \citep[although note that this might not always be true for a PWN, as it is believed to be the case for Vela-X, e.g.,][]{2013ApJ...774..110G}. The role of SNR G12.82-0.02 in the production of the TeV gamma-rays has been explored before by \cite{2007A&A...470..249F}, who found that both leptonic and hadronic scenarios are compatible with the multiwavelength observations of this object. If the GeV emission that we found, on the other hand, is produced by cosmic rays accelerated in this SNR, these particles would have already escaped from the object and would be propagating in the interstellar medium, as the SNR itself has a much smaller angular size ($\sim 3'$) compared to the $\sim 1.2\degr$ of the GeV emission found here.

Interestingly, there is a giant star forming region (SFR) known as W33 within the LAT source. It is characterised by having very massive stars and vigorous star formation activity \citep[e.g.,][]{2015ApJ...805..110M}. It subtends $15'$ and is located at a distance of 2.4 kpc \citep{2013A&A...553A.117I}. It has been proposed that star forming regions might be able to accelerate cosmic rays through the combined winds of massive stars and some of these regions have been detected in gamma-rays \citep[e.g.,][]{1983SSRv...36..173C,2001AstL...27..625B,2017ApJ...839..129K}, being the one in the Cygnus X region perhaps the most famous example \citep{2011Sci...334.1103A}. Known as the Cygnus Cocoon, the GeV emission is very extended ($\sim 4\degr$) and has a relatively hard spectrum (photon index $\sim 2.2$). Recently, \cite{2017ApJ...839..129K} found several extended regions of GeV emission within $\sim2\degr$ of the Galactic coordinates $(l,b)=(25\degr,0\degr)$ having similar photon spectral indices of $2.1\pm0.2$ and no spectral curvature in the LAT range, as well as no clear counterparts at lower energies. They present a plausible scenario where this emission is from a previously unknown SFR supported by the confinement of the GeV emission by bubble-like structures seen at lower wavelengths and related to the putative OB association, as well as similarities with the Cygnus Cocoon. Preliminary results in the \emph{Fermi} High-Latitude Extended Source Catalog show three similar objects tentatively associated to SFRs \citep{2017arXiv170906213W}.

\begin{figure}[ht!]
\includegraphics[width=9cm,height=5.5cm]{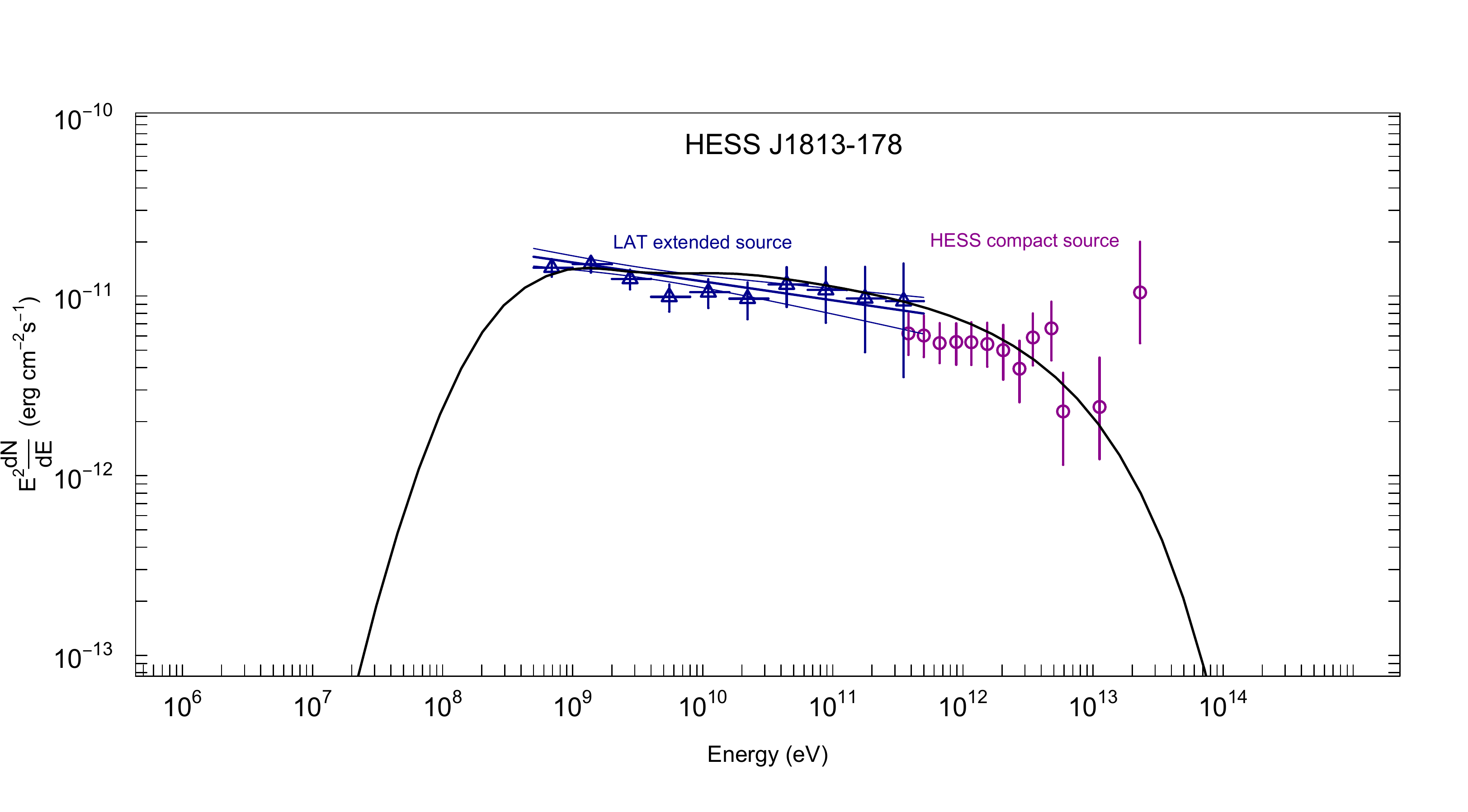}
\caption{Spectral energy distribution and hadronic model for the GeV source in the region of HESS J1813-178 (labeled ``LAT extended source''). The TeV data points for HESS J1813-178 are shown only for reference (labeled ``HESS compact source''). As explained in the text, since the TeV emission is compact and is believed to come from a PWN, while the GeV emission is considerably extended, the model shown here is not meant to apply to the TeV data points.}\label{fig5}
\end{figure}

Although the GeV emission found here is more extended than W33, perhaps this could similarly point to the presence of a previously unknown component of star formation which could be related to W33. The GeV spectral index, the extension and the lack of spectral curvature found here are features that are similar to those seen in the previous examples of SFRs. The diameter of the GeV region is $\sim 50$ pc for a distance of 2.4 kpc, about half the size of the Cygnus Cocoon. Following the usual motivation for hadronic models for the GeV emission from SFRs, due to their spectral shape, Fig. \ref{fig5} shows the gamma-ray SED of the extended emission in the direction of HESS J1813-178 with a hadronic model where the cosmic rays have a distribution in the form of a power-law with index 2.15 and a cutoff energy of 80 TeV (the TeV data points from H.E.S.S. are included only for reference, please note that the TeV emission is compact and possibly associated to a PWN). The required total energy in the particles is (normalized to the distance to W33) $$4.2\times 10^{50}\,\,\mbox{erg} \,\left(\frac{1\,\mbox{cm}^{-3}}{n}\right) \,\left(\frac{d}{2.4\,\,\mbox{kpc}}\right)^2,$$ which is similar to the energy found in the Cygnus Cocoon \citep{2011Sci...334.1103A}. We note that as found by \cite{2017ApJ...839..129K} a leptonic scenario for typical parameters in SFRs could also be used to explain the GeV gamma-rays (invoking bremsstrahlung emission), and we refer the reader to their work for an example of such a model.

Finally, other SNRs are known within the location of the GeV emission: G12.5+0.2, G12.7-0.0 and G13.5+0.2 \citep{2012AdSpR..49.1313F}. It is then possible that the extended gamma-rays are contributed by several of these objects which cannot be resolved by the LAT. A combination of accelerators may produce cosmic rays that escape and interact with molecular clouds. Some clouds are known to exist at least near HESS J1813-178 \citep{2007A&A...470..249F}. As mentioned before, rapid electron diffusion away from a PWN could produce an unusually steep GeV spectrum \citep{2011ApJ...743L...7H}. Even a previously unknown SNR located at a closer distance could be responsible for the emission and explain its extension. More observations are needed towards HESS J1813-178 to understand the environment and the origin of the high-energy particles.

\section{Summary}
Significant extended emission in the energy range 0.5-500 GeV was found with LAT data in the region of the TeV sources HESS J1809-193 and HESS J1813-178.

In the case of HESS J1809-193 the morphology of the GeV emission is compatible with the TeV morphology and the spectrum suggests a hadronic scenario, which is supported by other recent observations, however, mixed scenarios involving IC emission and hadronic emission as well as plausible bremsstrahlung fluxes at GeV energies cannot be discarded. In the hadronic scenario, the GeV-TeV data require cosmic rays with energies up to 1 PeV and, for a distance of 3 kpc to the source, an average target density for proton interactions of $\sim$10 cm$^{-3}$ is needed within the $\sim 1\degr$ region of high-energy emission for typical SNR shock energetics. This makes HESS J1809-193 a very interesting target for future TeV observations.

In the case of HESS J1813-178, the GeV emission is much more extended than the TeV emission ($\sim 1.2\degr$) and its spectrum is softer than that expected from the IC emission which is usually seen from leptons in a PWN (although an exception to this is seen in Vela-X). There is marginal evidence for a point source at GeV energies at the location of the peak of TeV emission, pressumably related to the PWN, besides the extended emission. The energetics, extended morphology and spectrum of the GeV emission are similar to those of SFRs that are thought to produce high energy particles (either leptons or hadrons), such as the Cygnus Cocoon. This region could be related to the W33 SFR but should have a larger angular size to explain the GeV emission. Other accelerators such as the SNRs that are known in the region could also contribute to the emission.

\acknowledgments

The comments made by the anonymous referee helped improve the quality of this work. Financial support from Universidad de Costa Rica and its physics department is acknowledged.

\bibliographystyle{aasjournal}
\bibliography{references}

\end{document}